\documentstyle[preprint,prl,aps]{revtex}

\begin{document}

\draft

\title{``Bare'' Effective Mass in Finite Sized $\nu = 1/2$ Systems} 

\author{A. Raghav Chari and F.D.M. Haldane}

\address{Department of Physics, Princeton University, 
Princeton, NJ 08544}

\maketitle

\begin{abstract}

In this note, we discuss the effective mass of quasiparticles in finite sized 
$\nu = 1/2$ systems in the lowest Landau level, given a natural notion we have 
of the Fermi surface in these finite sized systems. The effective mass is 
related to the difference between instantaneous density-density response 
functions of the ground state and excited states of the system; we do a finite 
size calculation for this difference for $\nu = 1/2$ systems with 7, 8 and 9 
fermions, and conjecture how the difference must look in the thermodynamic 
limit. 

\end{abstract}


The composite fermion ideas of Jain \cite{jain} lead us to believe that
we should see a metallic (compressible) state for fermions in the
half-filled Landau level. Halperin, Lee and Read \cite{HLR} and others
investigated the details of the physics of such a system analytically,
based around a mean-field Chern-Simons Landau-Ginzburg theory.
Recent numerical work has also shown \cite{HR} that there is a very
natural notion of the 
Fermi surface for $\nu = 1/2$ systems in the lowest Landau level. The basic 
entity in this picture is what is called a "cluster" wavefunction.

First pick $N$ (= number of fermions) momenta $\{ \vec{k}_{i} \}_{i = 1}^{N}$
such that they form a compact configuration in momentum space. The momenta will
have to be 
chosen consistent with the boundary conditions imposed on the system. In the 
thermodynamic limit, these boundary conditions will not be important and
we will 
be able to choose this cluster to be circular, but for finite system
sizes, the clusters can only be chosen to be approximately
circular. Then for any choice of $\{ \vec{k}_{i} \}_{i = 1}^{N}$ there
is a $\nu = 1/2$ LLL wavefunction defined as follows.  

The wavefunctions are built out of the $\nu = 1/2$ Laughlin state (which is 
bosonic because it is an even denominator state). In the standard holomorphic 
coordinates, this has the form

\begin{equation}
\Psi_{L}(\{ z_i \}) = \prod_{i<j} (z_i - z_j)^{2} \exp \left( -  
\frac{1}{4}|z_i|^2 \right)
\end{equation}

We will be doing finite sized calculations on the torus, and for this it is 
convenient to equivalently define the bosonic Laughlin state as the
ground state of bosons at $\nu = 1/2$ interacting with a hard-core
repulsive potential. Then in terms of this bosonic Laughlin state, we
can define a fermionic state for the cluster $\{ \vec{k}_i \}_{i = 1}^{N}$ as

\begin{equation}
\Psi ( \{ \vec{k}_i \}_{i = 1}^{N} ) = \det_{ij} \exp \left(
i\vec{k}_{i} \cdot \vec{R}_{j} \right) |\Psi_L (\nu = 1/2) \rangle
\label{slaterdet}
\end{equation}

\noindent
Note that this wavefunction has the correct statistics (fermionic) and filling 
factor. It has been shown elsewhere \cite{HR} that for any given set 
of periodic boundary conditions, the choice of $\{ \vec{k}_i \}_{i = 1}^{N}$ 
that gives the "most circular" compact configuration possible yields a 
wavefunction that is almost exactly the ground state for 
fermions interacting with coulomb interactions \cite{HR}. For example, the 
overlap between two arbitrary states is no more than a few percent, while the 
fermionic slater determinant state (equation~\ref{slaterdet}) for the "most 
circular" cluster has overlap better than 99 percent with the actual ground 
state for the system with coulomb interactions (this is true for related
short-range interactions as well). (We pick a 
natural notion of "most circular" by choosing the set $\{ \vec{k}_i 
\}_{i = 1}^{N}$ such that the quantity $\sum_{i \neq j} (\vec{k}_i - \vec{k}_j 
)^2$ is minimized, or equivalently, such that $\sum_{i} (\vec{k}_i
-\vec{k}_{\mathrm av})^2$ is minimized). We should also point out that
these cluster wavefunctions are {\em completely independent} of the
interaction between the fermions.

Given this result, and the definition for cluster wavefunctions 
(equation~\ref{slaterdet}), 
there is a very natural definition of the effective mass of
"quasiparticles" in this system.

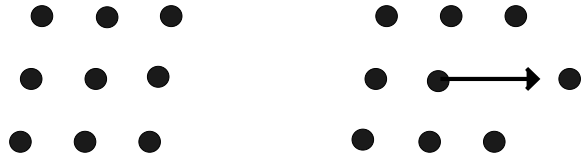
\begin{figure}
\caption{\label{cluster}
9 particle cluster in momentum space, shown here for a slightly oblique torus 
geometry. For this geometry, the cluster shown on the left is the "most 
circular" cluster we can exhibit for 9 particles. The cluster on the right 
depicts a basic particle-hole excitation of the "relaxed cluster on the left, 
and the arrow shows the fermi momentum, as per our operational definition in 
equation~\ref{fermimom}.}
\end{figure}

Let us denote the collection of momenta for the relaxed "most circular"
cluster by $\{ \vec{k}_{i}^{0} \}_{i=1}^{N}$ and let $\{ \vec{k}_{i}
\}_{i=1}^{N}$ depict an excited cluster (see figure~\ref{cluster} for an
example when $N = 9$).

Then the energy of the relaxed configuration is (let $H$ be the
Hamiltonian for fermions with coulomb interaction, for example)

\begin{equation}
E_0 = \langle \Psi(\{ \vec{k}_i^{0} \}) | H | \Psi(\{ \vec{k}_i^{0} \})
\rangle
\end{equation}

\noindent
which to a very good approximation is just the ground state energy for $\nu = 
1/2$ fermions with coulomb interaction. The energy of the excited cluster is

\begin{equation}
E = \langle \Psi(\{ \vec{k}_i \}) | H | \Psi(\{ \vec{k}_i \}) \rangle
\end{equation}

\noindent
In this setting, since the $|\Psi(\{ \vec{k}_i \}) \rangle$ are purely
functions of the $\{ \vec{k}_i \}$, so are the energies: $E = E( \{
\vec{k}_i \})$. The situation here is very similar to Hartree-Fock
theory, except here the Hamiltonian does not contain a contribution from
the kinetic energy (which gets quenched in the LLL). In the absence of
the kinetic energy, the energy functional $E(\{\vec{k}_i \})$ is translation
invariant; in addition, the energy ends up being a functional of the
occupation numbers $n_{\bf k}$ of the configuration, just as in the case
of Hartree-Fock (in Hartree-Fock theory,
however, the energy is quadratic in these occupation numbers while in
this case, the energy is {\em not} a quadratic functional of the occupation
numbers).

For simplicity, let us pick excited clusters that consist of a single 
quasiparticle-quasihole pair, or in other words, only one of the $\vec{k}_i$ 
differ from the $\vec{k}_i^{0}$ (See figure~\ref{cluster}). Label the
particles so that $\vec{k}_i = \vec{k}_{i}^{0}$ for $i = 2,...,n$. In
the thermodynamic limit, if $\delta \vec{k} \equiv \vec{k}_1 -
\vec{k}_1^{0}$ is small, we have

\begin{equation}
\delta E \equiv E - E_0 =  \frac{2 \delta \vec{k} \cdot \vec{k}_F}{m^{*}}
\label{deltae}
\end{equation}

\noindent
and (for $\nu = 1/2$) $k_F = l_0^{-1}$ 
where $l_0$ is the magnetic length. 
The definition yields what can be called the ``bare effective mass'' of
the composite fermion.
For finite sized systems, however, we need another operational
definition of the Fermi momentum. Define 
$\vec{k}_{\mathrm{av}} = \frac{1}{N} \sum_{i = 1}^{N}
\vec{k}_{i}^{0}$. Then we can compute 

\begin{equation}
\vec{k}_{F} = \left( \frac{\vec{k}_{1} + \vec{k}_{1}^{0}}{2} \right) - 
\vec{k}_{\mathrm{av}}
\label{fermimom}
\end{equation}

\noindent
For finite-sized systems, this definition of the Fermi momentum 
yields a vector of magnitude approximately equal to $1$. The reason we have to 
use a definition of this kind is that the clusters of momenta for the system 
sizes we examine are not rotationally invariant.

We can now reexpress the energy difference $\delta E$ in terms of the static 
structure factors of the two cluster wavefunctions. First, we notice
that in the lowest Landau level, we have

\begin{equation}
H = \sum_{q} V(q) \rho(q) \rho(-q)
\end{equation}

\noindent
where $\rho(q) = \sum_{i = 1}^{N} \exp(-i \vec{q} \cdot \vec{R}_{i})$.

\noindent
Thus

\begin{eqnarray}
E & = & \langle \Psi( \{\vec{k}_{i} \})| \sum_{q} V(q) \rho(q) \rho(-q)
| \Psi( \{\vec{k}_{i} \}) \rangle \nonumber \\
& = & \sum_{q} V(q) \langle \Psi( \{\vec{k}_{i} \})| \rho(q) \rho(-q) | \Psi( 
\{\vec{k}_{i} \}) \rangle  \nonumber \\
& = & \sum_{q} V(q) S[\Psi( \{\vec{k}_{i} \})](q)
\end{eqnarray}

\noindent
where $S[\Psi( \{\vec{k}_{i} \})](q)$ is the instantaneous structure
factor for the state $\Psi( \{\vec{k}_{i} \})$. Hence

\begin{equation}
\delta E = \sum_{q} V(q)( S[\Psi( \{\vec{k}_{i} \})](q) - S[\Psi( 
\{\vec{k}_{i}^{0} \})](q) )
\end{equation}

\noindent
Using equation~\ref{deltae} we get

\begin{equation}
\frac{1}{m^{*}} = \sum_{q} V(q) \left( \frac{S[\Psi( \{\vec{k}_{i} \})](q) - 
S[\Psi( \{\vec{k}_{i}^{0} \})](q)}{2 \delta \vec{k} \cdot \vec{k}_F} \right)
\label{effmass}
\end{equation}

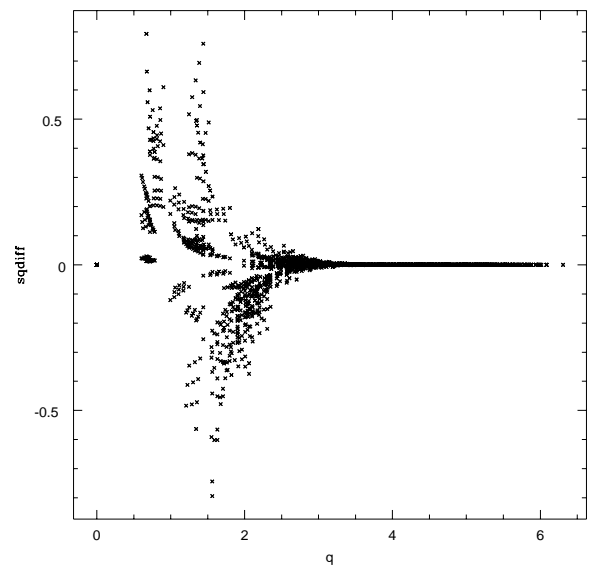
\begin{figure}
\caption{\label{seven}
$7$ fermion data for $\frac{S[\Psi( \{\vec{k}_{i} \})](q) - S[\Psi( 
\{\vec{k}_{i}^{0} 
\})](q)}{2 \delta \vec{k} \cdot \vec{k}_F}$. The relaxed cluster is the
same in all cases, though we have taken all possible excited clusters,
and plotted the cumulative data for a number of geometries.}
\end{figure}

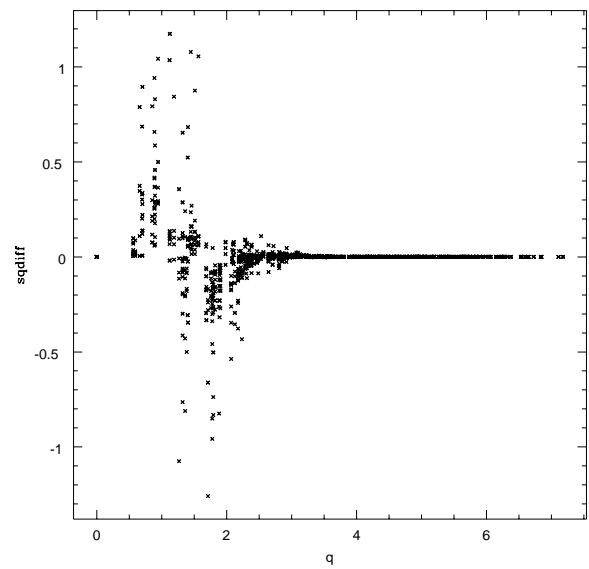
\begin{figure}
\caption{\label{eight}
$8$ fermion data for $\frac{S[\Psi( \{\vec{k}_{i} \})](q) - S[\Psi( 
\{\vec{k}_{i}^{0} 
\})](q)}{2 \delta \vec{k} \cdot \vec{k}_F}$. Same comments apply as for
7 fermion data.}
\end{figure}

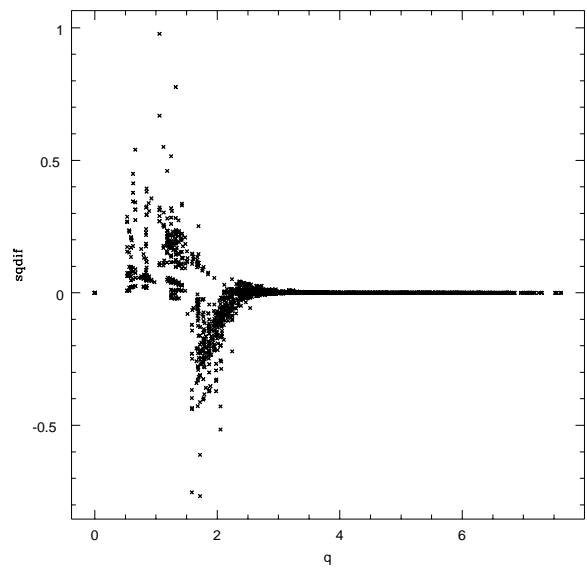
\begin{figure}
\caption{\label{nine}
$9$ fermion data for $\frac{S[\Psi( \{\vec{k}_{i} \})](q) - S[\Psi( 
\{\vec{k}_{i}^{0} 
\})](q)}{2 \delta \vec{k} \cdot \vec{k}_F}$. Same comments apply as for
7 fermion data.}
\end{figure}

We reiterate that the quantity in the large brackets in equation~\ref{effmass}
{\em depends only on the cluster wavefunctions, which are strictly
independent of the interaction}; all the interaction dependence of the
effective mass in this calculation follows rather simply from
equation~\ref{effmass} once we have calculated the relevant
(interaction-independent) static structure factors. Clearly, $1/m^{*}$
is a linear functional of $V(q)$.
In the thermodynamic limit, we should get the same answer for $1/m^{*}$ 
regardless of what particle-hole pair we create, just so long as the pair is 
created close to the Fermi surface. However, in finite sized systems, we 
cannot get arbitrarily close to the Fermi surface (defined by our relaxed 
cluster). Moreover, the relaxed cluster is not really circular, so
depending on what the choice of particle-hole pair we make, we will not
necessarily get the same result for $1/m^{*}$ each time. In 
figures~\ref{seven},~\ref{eight},~\ref{nine} we have shown the results of the 
calculation of $ \frac{S[\Psi( \{\vec{k}_{i} \})](q) - S[\Psi(
\{\vec{k}_{i}^{0} \})](q)}{2 \delta \vec{k} \cdot \vec{k}_F} $ for 7, 8
and 9 fermions at $\nu = 1/2$.

For each system size, we can vary the geometry (in other words, the angle 
between the sides of the box and the ratio of their lengths). Doing so allows 
us to probe a larger set of momenta than any single geometry will allow us to 
do, and in the process we hope to get a better handle on the physics in the 
thermodynamic limit. In the process of changing the geometry, the shape of the 
"relaxed" cluster $\{ \vec{k}_{i}^{0} \}$ will also change. In fact, if we 
change the geometry enough, we will encounter a phase transition to another 
region where the most relaxed cluster has a different set of momenta. In 
collecting the data for the figures~\ref{seven},~\ref{eight},~\ref{nine}, we 
picked only those geometries whose most relaxed cluster looked reasonably 
circular. In addition, in order not to have to deal with excitations
transverse to the Fermi surface, we threw out the data from all excited
clusters if the angle between $\delta \vec{k}$ and $\vec{k}_F$ exceeded
60 degrees. The choices made above are somewhat arbitrary, but are
reflective of the usual definition of the effective mass in the
thermodynamic limit.

The fact that we have finite sized systems produces the sort of spread in the 
data in figures~\ref{seven},~\ref{eight},~\ref{nine}. Essentially, all the 
different calculations for $1/m^{*}$ yield slightly different answers, but the 
qualitative features in the results for $\frac{S[\Psi( \{\vec{k}_{i} \})](q) - 
S[\Psi( \{\vec{k}_{i}^{0} \})](q)}{2 \delta \vec{k} \cdot \vec{k}_F}$
all follow the same pattern: the plot has a positive peak for some value
of $q$, and then for a larger value of $q$, there is a negative peak. We
can provide a heuristic argument for why the data looks this way, and
moreover, suggest what it must look lie in the thermodynamic limit. 

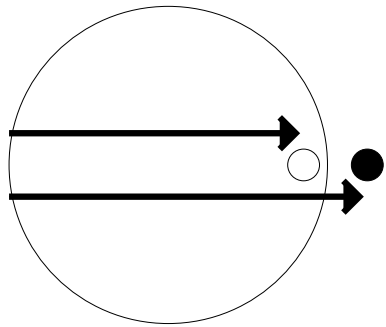
\begin{figure}
\caption{\label{heuristic}
The particle-hole pair shown suppresses processes contributing to $S(q)$
for the excited cluster relative to the relaxed cluster for $q$ slightly
greater than $2k_F$ and enhances processes for $q$ slightly less than $2k_F$.}
\end{figure}

Figure~\ref{heuristic} shows that the presence of a particle lying just
outside the Fermi surface (and a hole just inside) will suppress  the
matrix elements of $\rho(q) \rho(-q)$ for $q$ a little greater than $2
k_F$ and enhance it for $q$ a little less than $2 k_F$. This is in the
thermodynamic limit, but for our finite systems, $k_F$ is actually of
the same order as $| \delta \vec{k}|$ (see figure~\ref{cluster}). As a
result we see a positive peak at $q$ substantially less than $2k_F$ and 
a negative peak for $q$ somewhere around $2k_F$. It seems clear that if we had 
the ability to examine large enough systems, the data from all the different 
clusters ought to collapse onto a single curve, and moreover, the curve should 
consist of a very strong positive peak just under $2k_F$ and a very strong 
negative peak just over $2k_F$.

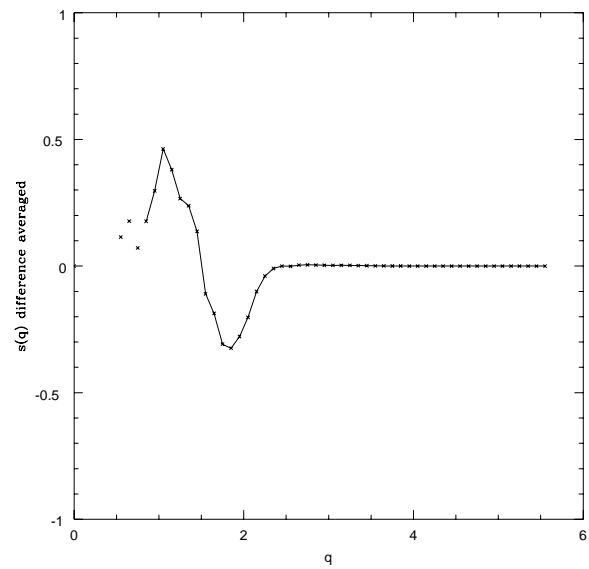
\begin{figure}
\caption{\label{average}
9 fermion data from figure~\ref{nine} averaged (and convolved with a
gaussian to  obtain a smoother plot). There is a small bump for small
$q$ that is an artifact of the finite size of the system, corresponding
to processes along the smallest possible momenta, after which there are
no processes until almost twice that $q$ value.}
\end{figure}

We can get an inkling for what this curve looks like by averaging out the data 
for the different clusters for a given system size, and this is shown in 
figure~\ref{average}. 

This is exactly what the heuristic described by figure~\ref{heuristic} would 
lead us to expect. The data we obtain can be used with
equation~\ref{effmass} to obtain the effective mass of quasiparticles in
this theory, and for coulomb interactions we end up with
$\frac{1}{m^{*}} \sim C \frac{e^{2}l_{0}}{\epsilon}$ where $C$ is a
number in the range $0.2 - 0.4$ depending on the geometry and choice of
excited cluster. This is in agreement with estimates made by others
\cite{HLR,SM}. For
arbitrary interactions, we can make some useful statements about the
form that $\frac{S[\Psi( \{\vec{k}_{i} \})](q) - S[\Psi(
\{\vec{k}_{i}^{0} \})](q)}{2 \delta \vec{k} \cdot \vec{k}_F}$ must take
in the thermodynamic limit. In the lowest Landau level, we can describe
the interaction in terms of Haldane pseudopotentials. On account of the
symmetry properties of the Laguerre polynomials that these
psedopotentials are defined in terms of, only the odd pseudopotentials
matter in the case of fermions. In other words, in the thermodynamic
limit, an exact curve (in place of figure~\ref{average}) multiplied by
$V(q)$ and integrated with respect to $2 \pi qdq$ should receive
contributions only from the odd pseudopotentials. For example, a
constant potential cannot make any contribution to the effective mass,
since a constant potential is described by the zeroth Laguerre polynomial.
We should stress that this is an important requirement for any formula for the 
effective mass of quasiparticles in the theory, when examined strictly within 
the lowest Landau level. 

We also note that Jain and Kamilla have recently \cite{JK} shown how to
evaluate the variational energy of composite fermion wavefunctions on
the sphere for much larger system sizes ($\sim$ 40-50 CFs). This they
achieve by modifying the definition of the CF wavefunction from the
original one proposed by Jain \cite{jain}. The properties of these
modified wavefunctions are apparently very similar to the original CF
wavefunction \cite{JK}. The $\nu = 1/2$ wavefunctions in the periodic
geometry can also be modified along the lines of \cite{JK}, potentially
allowing the calculation described here to be carried out on much larger
systems.

Finally, let us reiterate that within the given framework, there is no
way we will observe the kind of effective mass divergence
Halperin, Lee and Read calculate \cite{HLR} (which seems to depend
strongly on the
type of interaction between the fermions); this is because the ``bare''
effective mass we calculate is based on the static structure factor of
the cluster wavefunction, which is interaction independent. The
interaction dependence of the effective mass then follows
straightforwardly via equation~\ref{effmass}.

This work was supported by NSF DMR--9400362.

\end{document}